\pdfoutput=1

\documentclass[11pt]{article}

\usepackage{EMNLP2023}

\usepackage{times}
\usepackage{latexsym}

\usepackage[T1]{fontenc}

\usepackage[utf8]{inputenc}

\usepackage{microtype}

\usepackage{inconsolata}
\usepackage{graphicx}
\usepackage{booktabs} 

\usepackage{amsmath,amsthm,amsfonts,amssymb}
\usepackage{bm}
\usepackage{graphicx}
\usepackage{sidecap}
\usepackage{subcaption, wrapfig}
\usepackage{mathrsfs}
\usepackage{multirow, multicol}
\usepackage{parskip}  

\newcommand{\RN}[1]{%
	\textup{\lowercase\expandafter{\it \romannumeral#1}}%
}

\newtheorem{proposition}{Proposition}[section]

\newcommand{\calL}{\mathcal{L}}

\newcommand{\calT}{\mathcal{T}}

\def\1{\bm{1}}

\newcommand{\Kmat}[0]{{{\bf K}}}
\newcommand{\Vmat}[0]{{{\bf V}}}
\newcommand{\Qmat}[0]{{{\bf Q}}}
\newcommand{\Wmat}[0]{{{\bf W}}}

\DeclareMathAlphabet{\mathsfit}{\encodingdefault}{\sfdefault}{m}{sl}
\SetMathAlphabet{\mathsfit}{bold}{\encodingdefault}{\sfdefault}{bx}{n}

\def\hv{{\boldsymbol{h}}}

\def\xv{{\boldsymbol{x}}}
\def\yv{{\boldsymbol{y}}}
\def\zv{{\boldsymbol{z}}}
\def\tv{{\boldsymbol{t}}}
\def\mv{{\boldsymbol{m}}}
\def\thetav{{\boldsymbol{\theta}}}

\newcommand{\R}{\mathbb{R}}

\title{ALCAP: Alignment-Augmented Music Captioner}

\author{
Zihao He$^1$, Weituo Hao$^2$, Wei-Tsung Lu$^2$, Changyou Chen$^3$\\
\textbf{Kristina Lerman$^1$, Xuchen Song$^2$} \\
$^1$USC Information Sciences Institute; $^2$TikTok Inc.\\
$^3$Department of Computer Science and Engineering, University of Buffalo\\
\texttt{zihaoh@usc.edu}, \texttt{changyou@buffalo.edu}, \texttt{lerman@isi.edu}\\
\texttt{\{weituohao, weitsung.lu, xuchen.song@bytedance.com\}@bytedance.com} \\
}

\begin{document}
\maketitle

\def\thefootnote{\textbf{*}}\footnotetext{Work done when Zihao He interned at TikTok Inc.}\def\thefootnote{\arabic{footnote}}

\begin{abstract}
Music captioning has gained significant attention in the wake of the rising prominence of streaming media platforms. Traditional approaches often prioritize either the audio or lyrics aspect of the music, inadvertently ignoring the intricate interplay between the two. However, a comprehensive understanding of music necessitates the integration of both these elements. In this study, we delve into this overlooked realm by introducing a method to systematically learn multimodal alignment between audio and lyrics through contrastive learning. This not only recognizes and emphasizes the synergy between audio and lyrics but also paves the way for models to achieve deeper cross-modal coherence, thereby producing high-quality captions. We provide both theoretical and empirical results demonstrating the advantage of the proposed method, which achieves new state-of-the-art on two music captioning datasets. 
Our code is publicly available at \url{https://github.com/zihaohe123/ALCAP}.
\end{abstract}

\section{Introduction}
Learning to interpret music based on audio and lyrics has become an increasingly attractive research area for researchers in the filed music understanding and natural language processing \cite{manco2021muscaps,zhang2022interpreting}. The insights gained from this research into multimodal representation learning enable a wide range of applications such as streaming media discovery \cite{salha2021cold} and music recommendation with detailed and human-like descriptions \cite{andjelkovic2019moodplay}, making the dynamics of search and recommendation engines more explainable. However, captioning music is a challenging task. 
The duality of music, with its often nebulous and repetitious lyrics intertwined with intricate audio compositions harboring multiple layers of information, introduces a sophisticated web of complexities for models to navigate and understand.

Previous works on music captioning have primarily focused on refining singular facets of the encoder-decoder paradigm, from the enhancement of the music encoder to the incorporation of sophisticated attention mechanisms and beam search strategies. However, little effort has been directed towards leveraging the correspondence between audio and lyrics, which could potentially provide useful information for generating high-quality captions.
\citet{zhang2022interpreting} leverage the multimodal information from both lyrics and music through a cross-modal attention module, but the two modalities are not aligned before fusion. In reality, audio and lyrics are loosely aligned, as it is common for composers and lyricists to work separately in the music industry, resulting in different lyrics fitting the same melody. Additionally, the same words with different song patterns and styles can express diametrically opposite emotions. Therefore, the loose alignment between music and lyrics make them imperfect sources of data for existing multimodal learning methods that are not equipped with alignment mechanisms \cite{nichols2009relationships,zhang2022relyme}. In this regard, accurate and comprehensive music interpretation should leverage the subtle connections between music and lyrics.

In addressing the complexities of music understanding, we propose to align audio and lyrics pairs with contrastive learning before modality fusion and caption generation. Intuitively, paired audio and lyrics should be brought close together in the latent space, while non-paired ones should be pulled apart. By adding a contrastive loss, the multimodal input pairs are forced to be more aligned, which in turn guides the model to achieve stronger cross-modal consistency for a more meaningful fused latent space, thus generating better music captions.
To this end, we propose \textbf{Al}ignment Augmented Music \textbf{Cap}tioner (ALCAP), which is an extension of BART-fusion~\citep{zhang2022interpreting} for music captioning with a contrastive learning based alignment augmentation module. We provide a theoretical explanation of why the proposed alignment module results in improved generalization from an information bottleneck perspective. Extensive experiments on the Song Interpretation Dataset \cite{zhang2022interpreting} and the NetEase Cloud Music Review Dataset demonstrate ALCAP's superiority with marked improvements in key metrics (ROUGE-L and METEOR) over previous benchmarks.
We also observe performance gain of ALCAP in cross-modal text-music retrieval, which is a common application in industry, providing an indirect perspective to evaluate the caption quality. Lastly, we explore the effect of contrastive loss weights on the model performance via grid search and conclude our ablation study by showing that our proposed multimodal alignment module leads to more concentrated attention on language tokens through visualization analysis.


Our contributions are summarized as follows:
\begin{itemize}
    \item To the best of our knowledge, we are the first to propose an alignment augmentation module through cross-modal contrastive learning between music and lyrics for music captioning. By learning the interactions between the two modalities in an unsupervised manner, the model is guided to learn better cross-modal attention weights for meaningful fused latent space, leading to high-quality music interpretation generations.
    \item We provide a theoretical justification for the improved generalization of the proposed multimodal alignment module from an information theory perspective.
    \item Extensive experiments on two music captioning datasets demonstrate the effectiveness of our proposed alignment augmentation module, and we set the new state-of-the-art on the Song Interpretation Dataset. 
\end{itemize}

\section{Related Work}
\label{sec:related_work}

\subsection{Multimodal Alignment}
Multimodal representation learning has been increasingly important as modern intelligent applications require a comprehensive understanding of vision, language and speech \cite{yin2022x}. To learn meaningful latent spaces, unsupervised alignment between different modality inputs has been proven effective as an additional layer of structural information about the data. In the work of pretraining for speech synthesis \cite{bai20223}, aligning the acoustic and phoneme inputs makes the model more capable of learning cross-modal attention weights, thereby improving the quality of acoustic signal reconstruction. ALBEF \cite{li2021align} proposes to align vision and language before the modality fusion, purifying the multimodal input pairs, thus resulting in a more grounded vision and language representation. This approach can be interpreted as maximizing mutual information among different views of the same vision and language pair. $\mu$-VLA \cite{zhou2022unsupervised} introduces image-text level and region-phrase level alignment in vision and language pretraining so as to make the most of unpaired data. \citet{goyal2022retrieval} propose a retrieval process operating on past experiences to provide the agent with contextual relevant information, improving sample efficiency and representation learning of the policy function. It proves the effectiveness of retrieval-augmented module in continuous decision making process which also applies to the sequence of words generation \cite{ren2017deep,guo2018improving,yu2022coca,humphreys2022large}.

\textbf{Challenges in music-language alignment.}
While the majority of work in this field focuses on the alignment of vision and language \cite{radford2021learning}, the complexities of aligning audio, especially music, with text present a set of challenges fundamentally different from those of image-text alignment, as illustrated by the following reasons.
1) Richness of audio signals: Music, as an auditory medium, encompasses a rich variety of signals ranging from melodies, rhythms, timbres, to harmonic structures. These multifaceted signals, when combined, deliver a sonic experience that often possesses layers of meaning, emotional depth, and narrative nuances.
2) Ambiguity and subjectivity of lyrics: Lyrics, while being textual, are laden with poetic devices, metaphors, and often abstract representations. They can be open-ended, prompting multiple interpretations even without the musical context. When paired with music, lyrics can either complement the musical message or introduce added layers of ambiguity.
3) Synchronization challenges: Music and lyrics evolve synchronously over time. The alignment is not just about mapping the overall theme of a song to its lyrics; it is also about understanding how specific musical passages correspond to specific lyrical segments, reflecting shifts in emotion, intensity, or narrative. These challenges make the alignment of music and lyrics a unique and intricate problem. Hence, while the underlying principle of using contrastive learning might resemble existing models applied in image captioning, the intricacies of our domain necessitate tailored approaches. The application of such techniques to music captioning is relatively nascent, and our work aims to pave the way for more explorations in this direction.

\textbf{Comparison between ALCAP and CLIP.} While both models employ contrastive learning for multimodal alignment, their objectives and applications are distinct. CLIP \cite{radford2021learning} is designed for image-text understanding and generation, leveraging a large dataset of images with their corresponding textual captions. In contrast, ALCAP focuses specifically on music captioning by aligning audio and lyrics pairs using cross-modal contrastive learning before modality fusion. This novel alignment augmentation module in ALCAP is tailored to address the unique challenges posed by the ambiguous and repetitive nature of lyrics, as well as the complexity of audio signals in music.

\subsection{Multimodal Music Captioning} 
Music captioning is a challenging task as it requires the model to not only comprehensively understand both music and corresponding lyrics but also to avoid overfitting on limited music-lyrics pairs due to copyright restrictions. 
MusCaps \cite{manco2021muscaps} firstly addresses the music captioning task from an audio captioning perspective, using a multimodal input encoder-decoder architecture based on LSTM~\citep{hochreiter1997long}. While MusCaps achieves a performance boost in caption generation, its predictive word sequence is limited to 20 tokens, which narrows down the approach's applicability, or at least not suitable for our long and human-like language composition scenario. One of the most relevant works to ours is BART-fusion \cite{zhang2022interpreting} which is built on top of BART \cite{lewis2020bart}, adding a music encoder and modality fusion module. 
However, BART-fusion fails to 
fully mine the relationship between the music and lyrics input data. Inspired by works from retrieval augmented representation learning, we propose to improve the generalization ability of BART-fusion by introducing music and lyrics alignment before modality fusion.

\section{Methodology}
\label{sec:method}

\begin{figure*}[ht]
    \centering
	\includegraphics[width=0.9\textwidth]{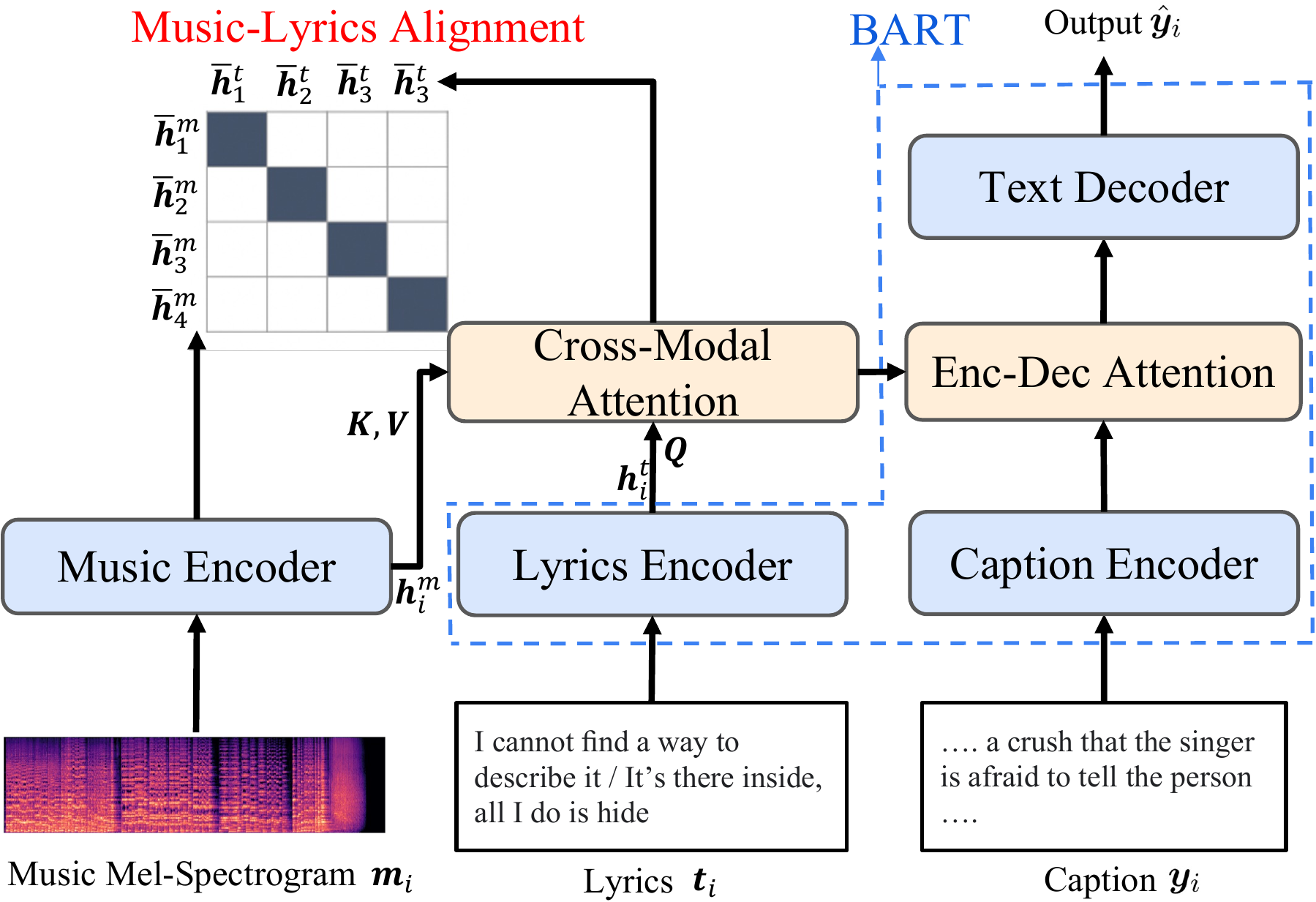}  
	\caption{An overview of ALCAP. The encoded representations of music and lyrics are first aligned using contrastive learning, then the aligned representations are fused using cross-attention, and further decoded through the text decoder. The architecture is based on BART \cite{lewis2020bart}.
 }
	\label{fig:overall_scheme}
\end{figure*}

In this section, we introduce the architecture of ALCAP, which is based on BART-fusion \citep{zhang2022interpreting}. 
We first state the problem definition, then go through each module of the architecture. The overall framework of ALCAP is shown in Figure \ref{fig:overall_scheme}.

\subsection{Problem Definition}
Given a song represented as a music-lyrics pair $\xv_i$, with a music track $\mv_i$ and its corresponding lyrics $\tv_i$, we aim to generate the caption (or interpretation) $\hat{\yv}_i$ of the song, consisting of a sequence of word tokens. In a typical setting of captioning, the attention-based encoder-decoder architecture is adopted to learn the mapping function from multimodal input to text output $f_{\thetav}: \{\mv_i, \tv_i\} \rightarrow  \hat{\yv}_i$. The model parameters $\thetav$ are optimized to generate the caption that is most consistent with the human annotated caption $\yv_i$.

\subsection{Multimodal Encoding}

\paragraph{Music Encoder} To obtain the representation of the music track, we use a pre-trained music encoder that includes a convolutional front-end and Transformer encoder layers \cite{won2019toward}. The model was originally trained to classify music audio into 50 tags under a multi-class setting using the Million Song Dataset \cite{bertin2011million}. These tags cover various musical characteristics, such as the genre (e.g., Jazz and Blues), mode, and the presence of specific instruments (e.g., piano or guitar). To perform the classification, the mel-spectrogram of a music track $\mv_i$ is first passed through a series of CNN layers for local feature aggregation in the time and frequency axis. The intermediate features are then fed into two Transformer encoder layers to model the information along the time axis, taking into account that elements of music can appear at different moments within a music clip. In \citet{won2019toward}, the output embedding series from the Transformer layer is further pooled to perform the classification task. However, in this paper, the embedding series $\hv^m_i \in \mathbb{R}^{l_m \times d_m}$ is used directly, where $l_m$ is the length of the music sequence and $d_m$ is the hidden dimension.

\paragraph{Lyrics Encoder} The representation of lyrics $\tv_i$ is obtained following standard BART encoder \citep{lewis2020bart}, and denoted as $\hv^t_i  \in \mathbb{R}^{l_t \times d_t} $, where $l_t$ is the length of the lyrics sequence and $d_t$ is the hidden dimension. The encoder consists of six multi-head self-attention layers.

\subsection{Multimodal Representation Alignment} 
Music and lyrics are not inherently connected, as different lyrics can fit the same melody, and the same lyrics can convey different emotions when paired with dynamic, rhythmic music. To fully capture the interactions between music and lyrics, we propose using contrastive learning before modality fusion to explicitly align the two modalities. This is expected to result in improved performance due to increased interactions between the two modalities, as has been previously shown to be effective in the vision and language domain \cite{li2021align}.

To be specific, given a batch of input music-lyrics pairs $\{(\mv_1, \tv_1)$, $(\mv_2, \tv_2)$, ....,$(\mv_n, \tv_n) \}$, we first obtain the music representations $\{ \hv^m_1$, $\hv^m_2$, ...., $\hv^m_n\}$ by the music encoder, and lyrics representations $\{ \hv^m_1$, $\hv^m_2$, ...., $\hv^m_n\}$ by the lyrics encoder respectively. As both music and lyrics are sequences, we denote $\bar{\hv}$ as the mean aggregation of $\hv$ along the sequence length dimension. Through a linear transform on $\bar{\hv}$, we obtain the latent code $\zv$ and use the InfoNCE loss \citep{oord2018representation} as the contrastive learning objective in latent space, as

\begin{align}\label{eq:sup_cl}
   \mathcal L_{contrast}= -\sum_{i=1}^{n} \text{log}\frac{\sigma(\zv^m_i \cdot \zv^t_i / \tau)}{\sum_{k }\sigma(\zv^m_i \cdot \zv^t_k / \tau)},
\end{align}
where $\zv^m_i$ and $\zv^t_i$ are the latent code of music and lyrics respectively, and $\sigma(\cdot)$ is the exponential function. For simplicity, we ignore the symmetric version by switching $\zv^m_i$ and $\zv^t_i$ in Equation~\ref{eq:sup_cl}, which is also applicable for the purpose of modality alignment. Note that InfoNCE can be interpreted as an estimator of a lower bound of mutual information~\cite{belghazi2018mutual,oord2018representation,cheng2020club}. We will incorporate this to prove the effectiveness of out proposed alignment module both theoretically and empirically, which is supposed to be non-trivial. We will revisit this in $\S$ \ref{subsec:mu} and $\S$ \ref{sec:exp}.

\subsection{Multimodal Fusion and Decoding}
Before decoding, the aligned representations of music tracks $\hv^m_i$ and lyrics $\hv^t_i$ are further fused by a cross-attention module, where the lyrics representations are linearly projected as queries, and the music representations are projected as keys and values. The process can be described as
\begin{align}
 &   \hv^{f}_i = \calT(\Qmat, \Kmat, \Vmat), \nonumber \\
& \Qmat = \Wmat^Q \hv^t_i,  \Kmat = \Wmat^{K} \hv^m_i,  \Vmat = \Wmat^V_{\ell} \hv^m_i,
\end{align}
where $\hv^{f}_i$ is the final fused representation, $\Wmat^Q \in \R^{d_t \times d_k}, \{\Wmat^{K}_{\ell}, \Wmat^V_{\ell}\} \in \R^{d_m \times d_k} $ are linear transform parameters, respectively; $d_k$ is the projection dimension.

The fused representation contains semantics from both the music track and the lyrics, as the alignment by contrastive learning ensures sufficient interactions between them. 
While the multimodal encoder fused the text and music as a whole, the decoding process follows a teacher-forcing fashion to predict each caption words, \emph{i.e.}, the ground-truth word token of the $i$th sample $\yv_{i,t}$ are provided at every step $t$ during training.
We use the BART decoder \citep{lewis2020bart} to generate the caption autoregressively and maximize the factorized conditional likelihood. The caption loss is defined as

\begin{equation}
    \calL_{\text{cap}} = -\frac{1}{n} \sum_{i=1}^n  \sum_{t=1}^{T}\text{log}P(\yv_{i,t}|\yv_{i,<t}, \hv_i^{f}),
\end{equation}

where $\yv_{i, <t}$ is the ground-truth word token before step $t$ and $P$ indicates the probability of the token at step $t$ conditioning on previous tokens and fused multimodal representation.

\subsection{Overall Learning Objective} To this end, we define  the final loss to be the weighted sum of the caption loss and the contrastive learning loss as follows: 
\begin{equation}
    \calL = \calL_{\text{cap}} + \alpha * \calL_{\text{contrast}},
\end{equation}
where $\alpha$ is the weight of the contrastive learning loss, balancing the contribution of captioning and multimodal alignment.

\section{An Information Theoretical Perspective}
\label{subsec:mu}
In this section, we explain the performance improvement of our alignment module based on contrastive learning from a mutual information perspective.

Given an input pair $\xv_i :=  \{\mv_i, \tv_i\}$, information bottleneck (IB) \citep{alemi2016deep} encourages the model to find minimal but sufficient information about the input $\xv_i$ with respect to the target caption words $\yv_i$. In other words, the objective of the training process in IB can be formulated as
\begin{align}\label{eq:ib}
    \max_{p_\theta(\zv|\xv)} \;   \text{I}(\yv;\zv) - \beta \text{I}(\xv;\zv),
\end{align} 
where $\text{I}(\yv; \zv)$ is the mutual information between the output and the latent code, $\text{I}(\xv; \zv)$ is the mutual information between the input and the latent code, and $p_\theta(\zv|\xv)$ is the conditional distribution of latent code parameterized by the encoder $\theta$. To optimize the IB, an upper bound on $\text{I}(\xv;\zv)$ is typically taken for generalization ability of a model \citep{tishby2000information,alemi2016deep}. 
From the information perspective, we show the following lower bound on the mutual information of $(\xv, \zv)$ in our setting. 

\begin{proposition}\label{prop:lowerbound}
The mutual information of $(\xv, \zv)$ in our setting is upper bounded by
\begin{align*}
    \text{I}(\xv;\zv) \leq \mathcal{R}(\zv) - \text{I}(m; t),
\end{align*}
where $\mathcal{R}(\zv) \triangleq \mathbb{E}_{p((\mv,\tv) | \zv)} \left[ \log \frac{\mathbb{E}_{p(\zv)}[p((\mv, \tv)|\zv)]}{p(\mv)p(\tv)} \right]$ depends only on $\zv$ and is independent of $\xv$.
\end{proposition}

In light of the fact that contrastive learning tends to maximize mutual information between $(\mv, \tv)$ pairs, the above lower bound suggests that it can be considered as an approximate implementation of information bottleneck. 
Furthermore, if the music-lyrics pairs used in contrastive learning are not well aligned, one can actually prove that the learning will fail.

\begin{proposition}\label{prop:randomcrl}
    If the music-lyrics pairing in the learning process is random such that the music and lyrics are sampled independently, then the mutual information between the input $\xv$ and the representation $\zv$ will be zero, and thus the encoder cannot learn anything useful.
\end{proposition}

The proof is provided in Appendix \ref{app:proof}. To sum up, based on the InfoNCE loss~\citep{gutmann2010noise}, the proposed alignment module can be interpreted as maximizing the mutual information lower bound between the music $\mv$ and corresponding text $\tv$, which translates to minimizing the mutual information between the input $\xv$ and the latent code $\zv$, and consequently improving the generalization ability of the model.

\section{Data}
In this paper, we experiment on two datasets -- the Song Interpretation Dataset \citep{zhang2022interpreting} and the NetEase Cloud Music Review Dataset.

\subsection{Song Interpretation Dataset}
The Song Interpretation (SI) Dataset dataset \cite{zhang2022interpreting} contains audio excerpts from 27,834 songs from Music4All Dataset \cite{santana2020music4all} and 490,000 user interpretations of the songs.
Each song is in 30 seconds and recorded at 44.1 kHz. Based on user votes of the interpretations, \citet{zhang2022interpreting} create three variants of the dataset, as 1) SI Full: the full dataset after some preprocessing; 2) SI w/voting $\geq$ 0: the subset with only interpretations that received non-negative votes;
3) SI w/voting $>$ 0: the subset with only interpretations that received positive votes.
The sizes of the training splits of the three datasets are 279,283, 265,360 and 49,736 respectively.
All three datasets share the same test split consisting of 800 instances.

\subsection{NetEase Cloud Music Review Dataset}
In addition to the Song Interpretation Dataset where the interpretations were mostly written by people who grew up under the influence of European and American culture, we curate another dataset - the \textbf{N}etEase \textbf{C}loud \textbf{M}usic (NCM) Review Dataset, where the reviews were written by people from China. NCM is a free music streaming service that is immensely popular in China. One of its most prominent features is that users can create their own playlists, write reviews and share the playlists with other users. 

We collect user-created playlists from NCM and keep those consisting of only English songs. 
Because our model generates captions at an individual song level, for each playlist, we keep one song from it that has the highest popularity, i.e., the song that has been collected to most playlists\footnote{Admittedly this is not the best way to create the song-review pairs given that the reviews were written at the entire playlist level. Nevertheless, the main goal of this paper is NOT to introduce this curated dataset, but to demonstrate the effectiveness of ALCAP in generating better song captions on different datasets.}. 
As a result, from each playlist, we have an instance of the song-review pair. 
For each song, we keep the middle 30 seconds excerpt and sample it at 22.05kHz. 
Since BART \cite{lewis2020bart} is pretrained in English, we translate the Chinese reviews into English using Google Translate.

\begin{figure}[ht]
    \centering
    \includegraphics[width=0.48\textwidth]{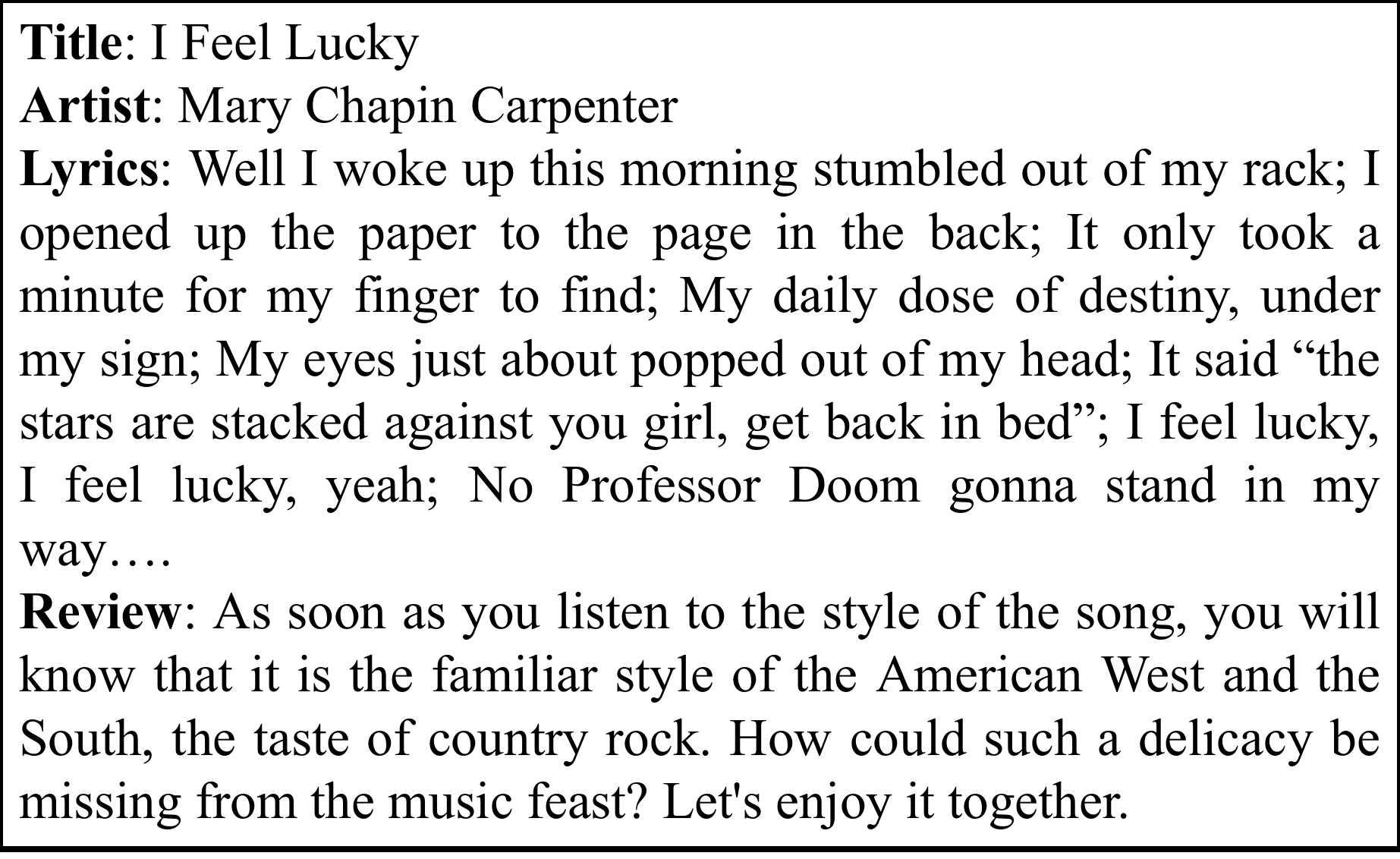}
    \caption{An example in NetEase Cloud Music (NCM) Review Dataset.}
    \label{fig:netease-example}
\end{figure}

The NCM Review Dataset contains 22,210 playlists (songs) and their reviews. An example is shown in Figure \ref{fig:netease-example}. We randomly split the dataset into train/val/test splits, with sizes of 15,547, 3,331, and 3,332.

\begin{table*}[ht]
\setlength{\tabcolsep}{12pt}
\def\arraystretch{1.05}
\centering
\addtolength{\tabcolsep}{-3pt}
\begin{tabular}{c||lcccc}
\toprule[1.2pt]
  Dataset                             & Method      & R-1   & R-2  & R-L   & Meteor  \\ \hline
                               & BART        & 44.1  & 14.0 & 24.5  & 22.5       \\
SI Full                       & BART-fusion & 46.1  & 15.0 & 25.1  & 23.0     \\
                               & ALCAP       & \textbf{48.2}$\pm$0.3  & \textbf{15.7}$\pm$0.1 & \textbf{26.4}$\pm$0.2  & \textbf{27.8}$\pm$0.2         \\ \hline
                               & BART        & 44.8  & 14.9 & 24.7  & 22.7      \\
SI w/voting $\geq$ 0 & BART-fusion & 46.7  & \textbf{15.6} & 25.5  & 23.4    \\
                               & ALCAP       & \textbf{47.7}$\pm$0.3   & \textbf{15.6}$\pm$0.1  & \textbf{26.0}$\pm$0.1   & \textbf{27.7}$\pm$0.1         \\ \hline
                               & BART        & 41.2  & 13.0 & 22.8  & 22.0     \\
SI w/voting > 0    & BART-fusion & 44.3  & 14.6 & 24.7  & 22.6    \\
                               & ALCAP       & \textbf{49.8}$\pm$0.1  & \textbf{16.0}$\pm$0.1 &  \textbf{27.1}$\pm$0.1 &  \textbf{27.7}$\pm$0.1      \\ \hline

\multirow{2}{*}{NCM Review}               & BART-fusion & 18.2$\pm$0.3 & 1.9$\pm$0.1 & 13.6$\pm$0.1 & 10.9$\pm$0.3     \\
                               & ALCAP       & \textbf{20.6}$\pm$0.2  & \textbf{2.6}$\pm$0.1 & \textbf{15.3}$\pm$0.3 & \textbf{11.8}$\pm$0.2   \\ 
\bottomrule[1.2pt]
\end{tabular}
\caption{Results of music captioning. The best results are highlighted in bold.}
\label{tab:results-music4all}
\addtolength{\tabcolsep}{3pt}
    
\end{table*}

\section{Experiments}
\label{sec:exp}

\subsection{Experimental Setup}
We resample each song at 16kHz and take a 15s excerpt. The maximum caption length is 512.
The model is implemented in PyTorch \citep{paszke2019pytorch}. We use the BART implementation \emph{facebook/bart-base} from Huggingface \citep{wolf2019huggingface}. We use a batch size of 26 and a learning rate of $5e-5$. The weight of contrastive learning $\alpha$ loss is set to 0.02. For better computation efficiency we freeze the parameters in the music encoder and precompute the music representations.
We train the model for 20 epochs and report the results on the test split using the checkpoint with the best evaluation performance. All hyperparameter tuning is based on grid search. All models are trained on a Tesla A100 GPU with 40GB memory. The training time for SI-Full, SI w/voting $\geq$ 0, SI w/voting $>$ 0, and NCM Review are 28h, 28h, 5h, and 3h respectively. 

We use ROUGE-{1,2,L} \citep{rouge2004package} and METEOR \cite{banerjee2005meteor} as evaluation metrics. ROUGE measures the overlap of n-grams between the referenced text and the generated text. On top of ROUGE, METEOR complementarily measures the semantic similarity between the two pieces of text by taking into account synonyms through WordNet. For both metrics, we use the implementation with default parameters from Huggingface Datasets library. We use three random seeds and report the average performance on the test set.

\subsection{Baselines}
BART is a model that utilizes only unimodal textual information from lyrics. The BART-fusion model, on the other hand, fuses representations from music and lyrics, but the two representations are not aligned prior to modality fusion. The results of these two baselines are reported in \citet{zhang2022interpreting}. We do not compare with \citet{manco2021muscaps}, which focuses on short-length music descriptions with a maximum of 22 tokens.

\begin{table*}[ht]

\setlength{\tabcolsep}{5pt}
\def\arraystretch{1.05}
\centering
\addtolength{\tabcolsep}{-2pt}
\begin{tabular}{llcccccccc}
\toprule[1.2pt]
Dataset                       & Method      & p@5            & p@10           & p@20           & p@30           & r@5             & r@10            & r@20            & r@30            \\ \hline
\multirow{2}{*}{SI Full}                      & BART-fusion & 3.2\%          & 1.9\%          & 1.2\% & 0.9\%          & 16.0\%          & 19.0\%          & 24.0\% & 27.0\%          \\
                              & ALCAP       & \textbf{3.6\%} & \textbf{2.1\%} & 1.2\% & \textbf{1.0\%} & \textbf{18.0\%} & \textbf{21.0\%} & 24.0\% & \textbf{31.0\%} \\ \hline
\multirow{2}{*}{SI w/voting \textgreater{}= 0} & BART-fusion & 2.2\%          & 2.0\%          & 1.0\%          & 0.7\%          & 11.0\%          & 17.0\%          & 20.0\%          & 23.0\%          \\
                              & ALCAP       & \textbf{4.2\%} & \textbf{2.6\%} & \textbf{1.5\%} & \textbf{1.1\%} & \textbf{21.0\%} & \textbf{26.0\%} & \textbf{30.0\%} & \textbf{32.0\%} \\ \hline
\multirow{2}{*}{SI w/voting \textgreater 0}    & BART-fusion & 2.2\%          & 1.2\%          & 0.9\%          & 0.7\%          & 11.0\%          & 12.0\%          & 18.0\%          & 20.0\%          \\
                              & ALCAP       & \textbf{3.0\%} & \textbf{1.6\%} & \textbf{1.0\%} & \textbf{0.8\%} & \textbf{15.0\%} & \textbf{16.0\%} & \textbf{20.0\%} & \textbf{23.0\%} \\ \hline

\multirow{2}{*}{NCM Review}                & BART-fusion & 0.2\% & 0.1\%          & 0.1\% & 0.1\% & 1.0\%  & 1.0\%           & 2.0\%           & 2.0\%           \\
                              & ALCAP       & 0.2\% & \textbf{0.2\%} & 0.1\% & 0.1\% & 1.0\%  & \textbf{2.0\%}  & \textbf{3.0\%}  & \textbf{4.0\%}  \\ 
\bottomrule[1.2pt]
\end{tabular}
\addtolength{\tabcolsep}{2pt}
\caption{Results of text-music retrieval. The best results are highlighted in bold.}
\label{tab:retrieval}
\end{table*}

\subsection{Experiments I: Music Captioning}
\label{sec:mus-cap}
The results are presented in Table \ref{tab:results-music4all}. 
We have found that ALCAP outperforms both BART-fusion and BART on all four datasets, in terms of all four metrics, thereby setting a new state-of-the-art. Specifically, the improvement on METEOR is more pronounced than on ROUGE metrics, which demonstrates that ALCAP is capable of capturing the semantics of the song for music captioning, not just memorizing the syntax. Furthermore, the results on the NCM Review for both models are overall worse than those on SI datasets. We believe this is due to the weaker correspondence between the music tracks and reviews in the NCM Review, as the reviews were originally created at the playlist level. Despite this, ALCAP is still able to capture such weak correspondence and achieve a significant improvement over the baseline.

\subsection{Experiments II: Text-Music Retrieval}

One of the most practical applications of music captioning is text-music retrieval, where given a piece of music description, the goal is to retrieve the most relevant music according to the text. In light of this, in this analysis, we test the retrieval capability of ALCAP and the baseline model. The setting of cross-modal retrieval in this experiment is different from previous works such as \citet{yu2022coca}, where the retrieval is performed on the two modalities that are directly aligned through contrastive learning.

As proposed in \citet{zhang2022interpreting}, we randomly select one sentence from the human-generated interpretation or review in the test split, and use it as a query. The queries are used to retrieve the corresponding songs through their generated captions by our models. Specifically, we compute the representations of the queries and generated captions using Sentence-BERT \cite{reimers2019sentence}. Thus, for each query, we obtain a ranked list of retrieved songs through the cosine similarities between the query representation and generated caption representations. We use precision@k and recall@k as the evaluation metrics. The results are shown in Table \ref{tab:retrieval}.

We observe that ALCAP outperforms BART-fusion on most datasets and metrics, indicating the superiority of cross-modal alignment between music tracks and lyrics that makes the generated captions more semantically aligned with human-written texts. This is apart from several cases where ALCAP ties with BART-fusion. Compared to SI datasets, the relatively low performance on NCM Review of both models is due to 1) the weak correspondence between the song and the review as we mentioned in previous sections, and 2) the retrieval pool (test split) is much larger -- 3,332 for NCM Review vs. 800 for SI. Nevertheless, ALCAP still outperforms the baseline in such a challenging scenario.

\subsection{Case Study I: Visualizing the Attention Weights}

\begin{figure}[ht]
    \centering
	\includegraphics[width=0.48\textwidth]{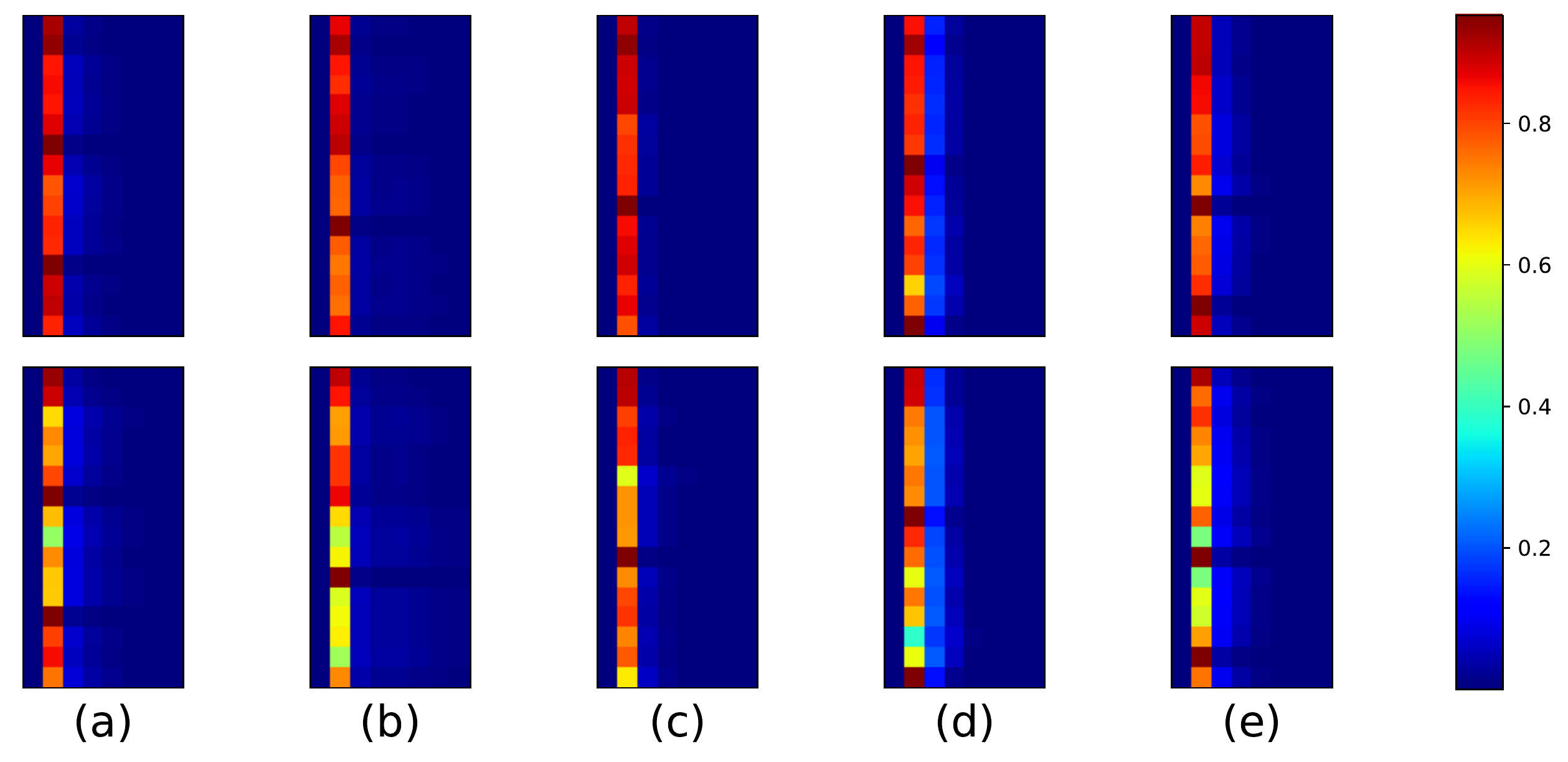}  
	\caption{Illustration of the cross-modal weights for five samples (a)$\sim$(e). The first row shows the cross-modal attention weights output by BART-fusion and the second row shows the weights by ALCAP. The y-axis and x-axis in each sub-graph indicates the text tokens and music segments respectively. 
 }\label{fig:attn_comp}
\end{figure}

To better understand the mechanism within the cross-attention module, we plot the attention weights of BART-fusion and ALCAP on five input examples from the training set in Figure \ref{fig:attn_comp}. Both models are trained on the SI w/voting $>$ 0 dataset.

The attention weights from ALCAP appear to be more focused on specific text tokens, in contrast to BART-Fusion, which has a more evenly distributed attention across all tokens. This phenomenon suggests that ALCAP, equipped with the cross-modal alignment module, is more effective at learning the interactions between the music audio and text domains.

\subsection{Case Study II: Examples of Generated Caption}
\begin{figure*}[!t]
    \centering
    \includegraphics[width=0.95\textwidth]{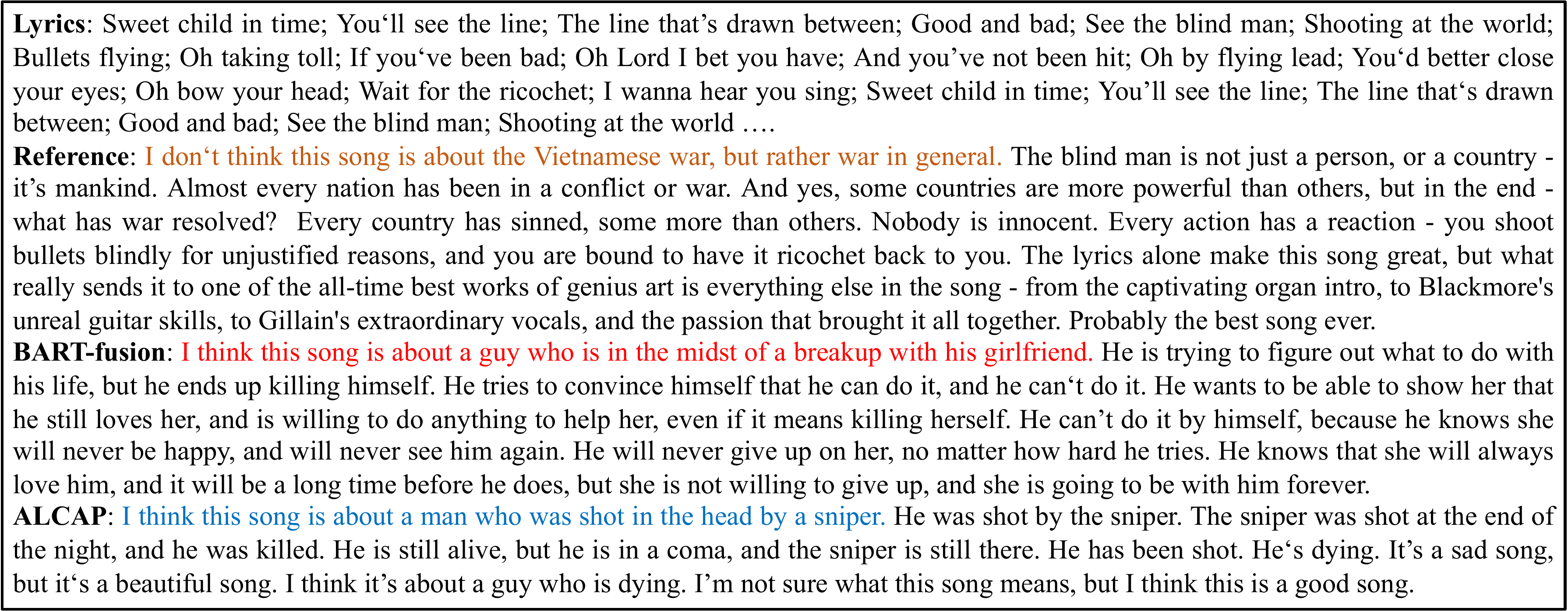}
    \caption{An example of generated captions from ALCAP and BART-fusion on \emph{Child In Time} by Deep Purple.}
    \label{fig:gen-example}
\end{figure*}

In this case study we show a representative example of generated captions from ALCAP and BART-fusion on \emph{Child In Time} by Deep Purple, as in Figure \ref{fig:gen-example}. The song is from the test split of SI, and both models are trained on SI w/voting $>$ 0.

From the lyrics and the reference interpretation, we can infer that the song is about war, which is captured by ALCAP. The generated caption contains "shot" and "sniper", which indicates that the model has correctly understood the theme of the song. However, BART-fusion fails to interpret the song correctly, instead interpreting it as a love song. We propose that this is due to the song's 70s Rock music style being too typical, and the lack of cross-modal alignment in BART-fusion. This allows the unimodal information from the sound track to dominate and confuse the model. As 70s Rock encompasses a wide range of topics, including love, it becomes harder to identify the correct topic of war. However, the alignment module in ALCAP manages to capture the semantics of the song and provide a more accurate interpretation.

\subsection{Ablation Study: Effect of Contrastive Learning Weight}

\begin{figure}[ht]
    \centering
    \includegraphics[width=0.48\textwidth]{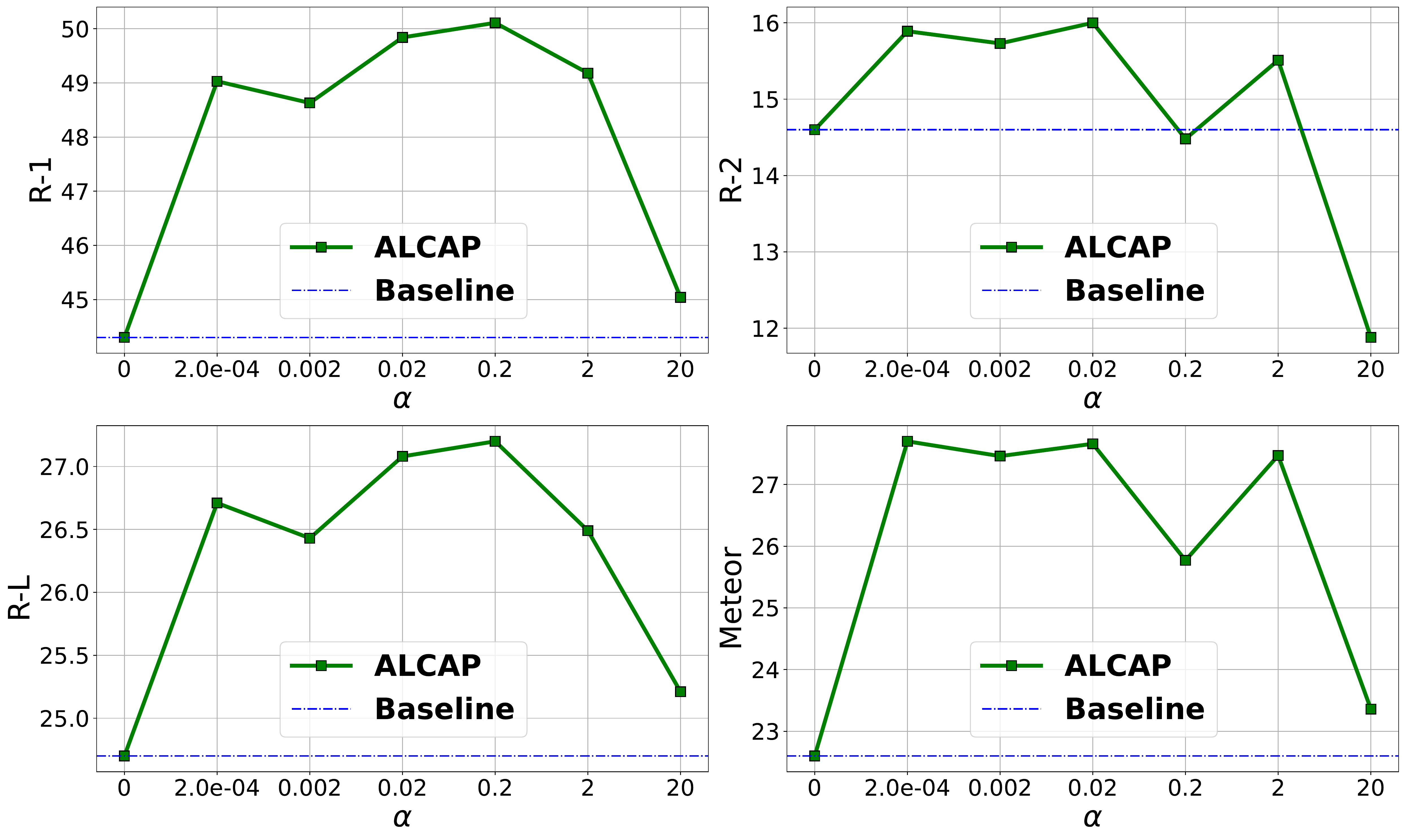}
    \caption{Results of music captioning using different weights of contrastive learning $\alpha$ on SI w/voting.}
    \label{fig:weights-caption}
\end{figure}

\begin{figure}[ht]
    \centering
    \includegraphics[width=0.48\textwidth]{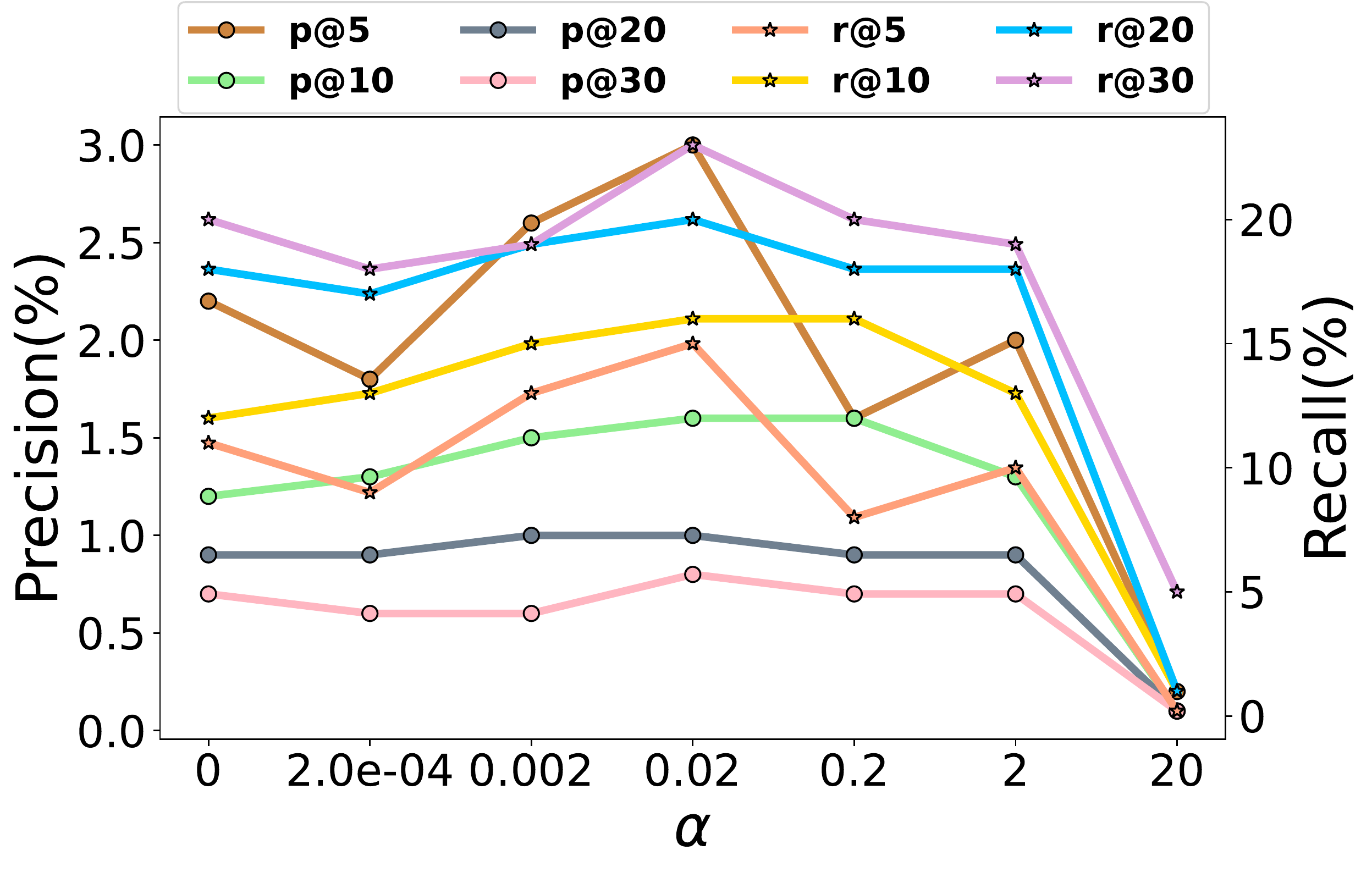}
    \caption{Results of text-music retrieval using different weights of contrastive learning $\alpha$ on SI w/voting.}
    \label{fig:weights-retrieval}
\end{figure}

To further investigate the effect of multimodal alignment through contrastive learning, we show the performances of using different weights of contrastive learning $\alpha$ on SI w/voting $>$ 0 on music captioning (Figure \ref{fig:weights-caption}) and text-music retrieval (Figure \ref{fig:weights-retrieval}). 

We observe that in both figures, the scores peak at $\alpha = 2e-2$, and decrease with higher weights or lower weights. When the weight is below $2e-2$, the model fails to learn sufficient alignment between the two modalities; on the other hand, when the weight is greater than $2e-2$, the model suffers because the overly large weight of contrastive learning loss negatively affects the optimizing of caption loss, which is the most prominent at $\alpha=20$.

\section{Conclusions}

In this paper, we introduce the \textbf{Al}ignment augmented music \textbf{Cap}tioner (ALCAP), a pioneering model designed to enhance music captioning by incorporating an alignment augmentation module using cross-modal contrastive learning. Our model's distinctiveness, particularly in the under-researched music domain, stems from its ability to successfully bridge the gap between music and its linguistic interpretation.
We provide a theoretical analysis of the improved generalization of our model from an information bottleneck perspective. Experiments on two music captioning datasets demonstrate the effectiveness of ALCAP, and we achieve the new state-of-the-art on both of them.

Our next steps will focus on collecting extensive music song data from the Web and pretrain the music-lyrics alignment module, after which the model is further finetuned on small-scale music captioning. This model would be tailored to cater to large-scale song interpretation generation on various genres and styles of music. In addition, ground-truth interpretations, as drawn from user-generated content, inherently bring the risk of biases. We will be committed to developing strategies to effectively mitigate biases from these interpretations, ensuring that our model's outputs are as neutral and accurate as possible.


\section*{Limitations}
Due to computational limitations, the parameters of the music encoder in ALCAP were fixed, and the music representations were precomputed, following \citet{zhang2022interpreting}. This approach may result in a decrease in performance compared to a model where the music encoder is fully fine-tuned for the music captioning task. Additionally, the Song Interpretation dataset, being the only publicly available music captioning dataset, is limited in scope, making it challenging to pretrain a large music captioning model that is suitable for various genres and styles of music. Furthermore, user-generated song interpretations and reviews may contain biases or even hate speech, which could be perpetuated during training of the model.

\section*{Acknowledgements}
Changyou Chen is partially supported by NSF AI Institute-2229873, NSF RI-2223292, an Amazon research award, and an Adobe gift fund. Any opinions, findings and conclusions or recommendations expressed in this material are those of the author(s) and do not necessarily reflect the views of the National Science Foundation, the Institute of Education Sciences, or the U.S. Department of Education.

\bibliography{anthology,custom}
\bibliographystyle{acl_natbib}

\appendix
\section{Proofs of Proposition 4.1 and 4.2}
\label{app:proof}

\begin{align*}
    \text{I}(\xv;\zv) = & \mathbb{E}_{p(\xv,\zv)} \left[ \log \frac{p(\xv, \zv)}{p(\xv)p(\zv)} \right] \nonumber \\
    = & \mathbb{E}_{p(\xv,\zv)} \left[ \log \frac{p(\xv|\zv)}{p(\xv)} \right] \\
    =& \mathbb{E}_{p(\mv,\tv, \zv)} \left[ \log \frac{p((\mv, \tv)|\zv)}{p(\mv, \tv)} \right]  \nonumber \\
    =& \mathbb{E}_{p(\mv,\tv, \zv)} \left[ \log \frac{p((\mv, \tv)|\zv)}{p(\mv)p(\tv)} \right] \\
    &\indent -I(m; t) \\
    =& \mathbb{E}_{p((\mv,\tv) | \zv)p(\zv)} \left[ \log \frac{p((\mv, \tv)|\zv)}{p(\mv)p(\tv)} \right] \\
    &\indent - I(m; t) \\
    \leq& \mathbb{E}_{p((\mv,\tv) | \zv)} \left[ \log \frac{\mathbb{E}_{p(\zv)}[p((\mv, \tv)|\zv)]}{p(\mv)p(\tv)} \right] \\
    & \indent - I(m; t) \\
    =& \mathbb{E}_{p((\mv,\tv) | \zv)} \left[ \log \frac{p(\mv, \tv)}{p(\mv)p(\tv)} \right] \\
    &\indent - I(m; t) ~,
\end{align*}
where the inequality follows by Jensen inequality. 
This completes the proof of Proposition~\ref{prop:lowerbound}.

Based on the above derivation, if $(\mv, \tv)$ pairs are sampled randomly, in the probabilistic graphical model language \cite{koller2009probabilistic}, this corresponds to a $V$-structure between $(\mv, \tv)$ and $\zv$. And a $V$-structure indicates the marginal independency between $\mv$ and $\tv$ \cite{koller2009probabilistic}. Thus, we have

\begin{align*}
    \text{I}(\xv;\zv) \leq &  \mathbb{E}_{p((\mv,\tv) | \zv)} \left[ \log \frac{p(\mv, \tv)}{p(\mv)p(\tv)} \right] - I(m; t) \\
    =& \mathbb{E}_{p((\mv,\tv) | \zv)} \left[ \log \frac{p(\mv)p(\tv)}{p(\mv)p(\tv)} \right] - I(m; t) \\
    &= -I(\mv; \tv)
\end{align*}

Since we know that both $I(\xv, \zv)$ and $I(\mv; \tv)$ must be non-negative, we have
\begin{align*}
    I(\xv; \zv) = I(\mv; \tv) = 0~.
\end{align*}
Consequently, this leads to the independency of $\xv$ and $\zv$, {\it i.e,}, $\zv$ contains zero information of $\zv$. 
This completes the proof of Proposition~\ref{prop:randomcrl}.

\end{document}